# Giant second harmonic generation in supertwisted WS₂ spirals grown in step edge particle induced non-Euclidean surfaces

Tong Tong[1,2,‡], Ruijie Chen[1,‡],Yuxuan Ke[3], Qian Wang[1], Xinchao Wang[1], Qinjun Sun[2], Jie Chen[1], Zhiyuan Gu[1], Ying Yu[1], Hongyan Wei[1], Yuying Hao[1], Xiaopeng Fan[1,*], and Qing Zhang[3,*]

[1]College of Electronic Information and Optical Engineering, Taiyuan University of Technology, Taiyuan 030024, China.

[2]College of Physics, Taiyuan University of Technology, Taiyuan 030024, China.

[3]School of Materials Science and Engineering, Peking University, Beijing 100871, China

Email: xiaopengfan@tyut.edu.cn; q_zhang@pku.edu.cn.

**Abstract**

In moiré crystals resulting from the stacking of twisted two-dimensional (2D) layered materials, a subtle adjustment in the twist angle surprisingly gives rise to a wide range of correlated optical and electrical properties. Herein, we report the synthesis of supertwisted $WS_2$ spirals and the observation of giant second harmonic generation (SHG) in these spirals. Supertwisted $WS_2$ spirals featuring different twist angles are synthesized on a Euclidean or step-edge particle-induced non-Euclidean surface using a carefully designed water-assisted chemical vapor deposition. We observed an oscillatory dependence of SHG intensity on layer number, attributed to atomically phase-matched nonlinear dipoles within layers of supertwisted spiral crystals where inversion symmetry is restored. Through an investigation into the twist angle evolution of SHG intensity, we discovered that the stacking model between layers plays a crucial role in determining the nonlinearity, and the SHG signals in supertwisted spirals exhibit enhancements by a factor of 2 to 136 when compared with the SHG of the single-layer structure. These findings provide an efficient method for the rational growth of 2D twisted structures and the implementation of twist angle adjustable endowing them great potential for exploring strong coupling correlation physics and applications in the field of twistronics.

**Introduction**

The emerging twisted van der Waals (vdW) structures offer a variety of attractive building blocks for the study of various extraordinary physical phenomena.错误!未找到引用源。·错误!未找到引用源。 For example, the twisted vdW structures with multiple vertically stacked layers create periodic moiré lattices due to lattice mismatches or twist angles.错误!未找到引用源。 The twist angle determines the dynamics within the mini-Brillouin zone,错误!未找到引用源。·错误!未找到引用源。 which induces strongly correlated quantum phenomena ranging from moiré excitons,错误!未找到引用源。·错误!未找到引用源。 moiré phonons 错误!未找到引用源。·错误!未找到引用源。 and enhanced photoresponse 错误!未找到引用源。·错误!未找到引用源。 in optics to superconductivity,错误!未找到引用源。 as well as ferromagnetism 错误!未找到引用源。 and fractional quantum Hall effects 错误!未找到引用源。·错误!未找到引用源。 in electronics. In particular, the twisted vdW structures feature an almost arbitrary geometry that is consistent with the crystallographic symmetry groups of the sublattices, and therefore offer an avenue for quantum manipulation of quasiparticles in quantum optics. Recently, the supertwisted spiral moiré lattices in supertwisted $WS_2/WSe_2$ spirals were resolved, enabling special twist angle-dependent morphologies.错误!未找到引用源。 In contrast to prior multilayer stacked moiré materials assembled through direct stacking of monolayers or multilayers, the spiral structure constitutes a continuous single-layer rotating stacking arrangement. This configuration not only facilitates a pathway for the transport of correlated quantum state electrons but also hold promise for exploring the strong coupling correlation physics of twistronics,

錯誤!未找到引用源。, 錯誤!未找到引用源。 including optics, electronics, acoustics, condensed matter and quantum physics.錯誤!未找到引用源。 Supertwisted spiral moiré lattices hold immense promise for relational quantum fundamental research and practical applications.[21] Meanwhile, the twist angle introduces a degree of freedom for intentionally manipulating the nonlinear geometric phase to realize phase compensation in twisted spiral structures, and improve the nonlinear conversion efficiency.[25] However, the relationship between the second harmonic generation (SHG) effect of 2D twisted vdW structures and the twist angle of the layers needs further investigation, particularly for 2D supertwisted spirals with more than two layers.

Here we investigate the evolution of the SHG effect based on supertwisted $WS_2$ spirals obtained on a Euclidean or step edge particle induced non-Euclidean surface. We present the experimental realization of supertwisted spiral with continuous twist angles and modified symmetry. The supertwisted $WS_2$ lattices are obtained by screw-dislocation-driven (SDD) growth and have a tunable twist angle. Depending on the twist angle, spiral structures with a different periodic structure enable inversion symmetry breaking and thus a giant SHG effect. Overall, these findings pave a path to manipulate the structural symmetry in vdW layered materials for further studies in strongly correlated physics and twistronics.

**Results and Discussion**

**One-step growth of step edge particles induced supertwisted $WS_2$ spirals**

vdW materials are materials with strong in-plane covalent bonding and weak interlayer interactions. The 2D nature of these materials makes them easy to bend and less sensitive to the epitaxial relationship of the underlaying substrate. Those properties make it easier for vdW materials growth be influenced by the geometry of the growth substrate. We realize the synthesis of supertwisted $WS_2$ spirals via a water vapor-assisted chemical vapor deposition (CVD) process (Fig. S1). Briefly, $WS_2$ powder was filled in a quartz boat and then placed in the heating center of the furnace, and a silicon substrate with 280 nm $SiO_2$ was placed downstream for material deposition (See Materials and Methods for growth details). The key step in the growth of supertwisted spiral $WS_2$ is the introduction of deionized water. In particular, 0.05–0.1 ml deionized water was dropped onto the edge of another quartz boat. This keeps the water droplets in an elliptical shape and prevents them from evaporation before heating. The boat with water was then placed 20–25 cm upstream from the $WS_2$ powder (outside the insulation of the tube furnace to prevent the deionized water from evaporating too quickly when heated). In previous work, we have demonstrated that the water here can serve as a transport agent to promote the dissociation and volatilization of the $WS_2$ precursor[26,错误!未找到引用源。7] by forming volatile $WO_x(OH)_y$ gas species and $H_2S$ gas. Those gas species lead to increased vapor pressure and, accordingly, increased nucleation. The volatile $WO_x(OH)_y$ can also decompose and form $WO_x$ particles, which tends to attach to the edges of the growing $WS_2$. Those $WO_x$ particles can act as the nucleation site to a screw dislocation (red dotted circle in Fig. 1a and inset model in Fig. 1c). As growth proceeds, the continuously supplied vapor species from the precursor enable additional nucleation and growth on the newly formed ribbons, ultimately forming the supertwisted $WS_2$ spirals (Fig. S2a-h).

Fig. 1a and 1b are the optical and atomic force microscopy (AFM) images of a left-handed supertwisted $WS_2$ spirals, which is located on top of a $WO_x$ particle formed on the step edge of $WS_2$ nanoribbon. AFM shows the height of the black band underneath of the $WS_2$ spiral is about 1.4 nm (Fig. 1c), which corresponds to the thickness of bilayer $WS_2$. It indicates that a bilayer $WS_2$ ribbon is underneath the $WS_2$ spiral. During the SDD growth process, the $WO_x$ particles are formed on the edge of the nanoribbons, as shown by the red dashed circle in Figure 1a. The formation of these particles provides a non-Euclidean substrate for the growth of the spiral structures (inset in Fig. 1c). Fig. 1d and 1e are the optical and AFM images of a right-handed supertwisted $WS_2$ spiral with ≈17° twist angle. The twist angle is defined by the acute angle between the layer edges of adjacent layers (Fig 1f). When the spiral grows on a cone surface, it will also be bent following the conic curvature with a consistent twist angle between each successive layer. This uneven strain could result in small variations of twist angles (approximately 0.5 degree) during the growth of spiral bending with the conic surface.[21,27] And the height of each layer of the $WS_2$ spiral is about 0.7 nm (Fig. 1g). Fig. S2f-h provides more details of the left/right-handed supertwisted $WS_2$ spirals. Optical microscope allows us to quickly and roughly identify the spirals with different twist angles. Under an optical microscope, the spirals with small twist angles have triangular shapes (Fig. S2a ,2c, 2f and 2g), while the spirals with large twist angles show a circular shape (Fig. S2d, 2e and 2h). Fig. 1h show the structural model of the left/right-handed supertwisted WS$_2$ spiral with different twist angle, triangular shapes or circular shape is a good indication to roughly estimate the twist angles of the supertwisted structures.

The shape of these supertwisted spirals can be explained by the geometric mismatch between lattice and the underlying substrate[21]. Because 2D materials are easy to bend and less sensitive to the epitaxial relationship of the underlaying substrate, they trend to follow the geometry of the underlaying substrate[28-30] despite inducing strain. In typical CVD of WS$_2$,错误!未找到引用源。,错误!未找到引用源。 the WS$_2$ spirals are grown on a flat SiO$_2$/Si substrate, which is a Euclidean surface with an angular period of $2\pi$. Therefore, the WS$_2$ spiral grows into a triangle with the edges of different layers aligned in parallel (Fig. 2a-c).[31,32] However, when the screw dislocation locates on a protrusion which is a non-Euclidean surface, the SDD growth would result in a very different morphology called supertwisted spirals (Fig. 2d-k).错误!未找到引用源。 There are two different cases when growing supertwisted spirals on a conical surface: (i) Fig. 2d shows the angular period of the substrate is $2\pi$-$\alpha$ ($\alpha$>0), which corresponds to a crystal grows at the apex of a cone. Thus, each complete lattice period ($2\pi$), the lattice carries an excess angle of $\alpha$ going into the next cycle (Fig. 2e). For a right-handed spiral (counterclockwise from bottom to top), this causes the triangular edges in each layer rotated counterclockwise by an angle of $\alpha$ with respect to the bottom layer, resulting in a right-handed supertwisted spirals, as shown in Fig. 2f. In this case, the center of the dislocation coincides with the top center of a protruding cone, as shown in Fig. 2g. (ii) Fig. 2h indicates that the angular period of the substrate surface is $2\pi+\alpha$ ($\alpha>0$), which corresponds to a warped "hyperbolic cone" surface. Since the angular period of the surface is larger than the lattice period, one lattice period is unable to cover the surface of the substrate (Fig. 2i). As a result, an angle of $\alpha$ needs to be "borrowed" from the next lattice period. This causes the triangular edges in each layer rotated clockwise by an angle of $\alpha$ with respect to the previous layer, resulting in left-handed twisted spirals, as shown in Fig. 2j. In this case, the center of the dislocation coincides with the edge of the protruding cone (Fig. 2k). The key factor enabling the controlled preparation of supertwisted spiral materials with specific twist angles involves combining the SDD growth mechanism with non-Euclidean substrates (refer to Fig. 2g and 2k). As depicted in Fig. 2f and 2j, with each period, the subsequent edge rotates by an angle $\alpha$ relative to the edge in the previous period. Achieving precise control over the twist angle hinges heavily on the relative alignment between the spiral axis and the conical surface. When the spiral structure's axis is located on the cone surface (Fig. 2g), the twist angle $\alpha$ (Fig. 2f) is dictated by the slope of the cone surface (as shown in Fig. 2d and 2e). A steeper inclination of the cone surface facilitates the production of supertwisted spiral materials with larger twist angles.[27] If the spiral axis lies outside the cone surface (as depicted in Fig. 2k), the slope of the cone surface solely impacts the twist angle. As the distance between the spiral axis and the cone surface increases, the influence of the cone surface's slope on the twist angle diminishes until it becomes negligible. Eventually, this results in the formation of a spiral with a 0° twist angle, as illustrated in Fig. 2c.

Furthermore, scanning transmission electron microscopy (STEM) was conducted on the spiral $WS_2$ structure to validate the precision of determining the twist angle along the edges. Low-resolution high-angle annular dark-field (HAADF) imaging of the partial structure reveals the layer edges (see Fig. 3a), with different colored lines representing edges of various layers within the spiral structure. Specifically, red, blue, green, and purple denote single-, double-, triple-, and quadruple-layer regions, respectively. Fig. 3b displays the outcome of position-averaged convergent beam electron diffraction (PACBED) pattern acquired from the purple mark in Fig. 3a, clearly illustrating consistent twist angles between adjacent layers, indicated by the white arrow in Fig. 3b. The STEM analysis further corroborated the accuracy of the twist angle determination along the edges,[21] as shown in Fig 3c and 3d. Owing to the triple symmetry of layered $WS_2$ materials, the twist angle between successive layers in the spiral structure ranges from 0° to 30°. We chose the twisted spirals with the twist angle of 0°, 13°, 16° and 20° to perform nonlinear optical studies (Fig. 3e-l). We refrained from choosing spirals with very small twist angles because our optical spectrum system lacked the capability to spatially distinguish individual layers within these spirals (Fig S3 and Fig S4). Similarly, we didn't investigate spirals with large twist angles due to the substantial inversion symmetry changes induced by such angles.[31]

**Twist angle-dependent nonlinear optical effects**

We examined the nonlinear optical properties of these atomically $WS_2$ spirals. The relationship between the twist angle-dependent SHG and the thickness-dependent SHG of $WS_2$ spirals were investigated using a reflection mode. The thickness-dependent SHG intensity of supertwisted $WS_2$ spirals shows that all regions have a remarkable SHG signal (Fig. 4).

Fig. 4a show the SHG spectra obtained from a normal triangular $WS_2$ spiral ($\approx 0°$ twist angle, Fig. S3). Its SHG intensity increases with increasing layer numbers, and the maximum SHG signal in 30L is enhanced over 136 times compared to monolayer $WS_2$ (Fig. S5a, b). This is because this $WS_2$ spiral has the 3R phase structure with broken inversion symmetry.[31,35] As the layer numbers increase, the SHG intensity increases rapidly, giant SHG was therefore observed in these triangular $WS_2$ spirals (Fig. 4e). However, when $WS_2$ spiral is a supertwisted structure with non-zero twist angle (Fig. 4b-c), there's an oscillating SHG intensity (Fig. 4f-h) with increasing layer numbers, and a larger twist angle of supertwisted spiral leads to the presentation of the SHG intensity with larger oscillation frequency. Yet, the SHG intensity of the supertwisted spirals is always higher than that of monolayer $WS_2$. For example, in the supertwisted $WS_2$ spiral with a twist angle of 13° (Fig. 4b and 4f), the SHG intensity in 1L–5L gradually increases from 4 times to 92 times as layer number increases compared to monolayer $WS_2$, just like the $WS_2$ spiral with 3R phase. However, the SHG intensity decreased rapidly in 6L-7L, the SHG intensity is still higher than that of monolayer $WS_2$. The SHG intensity of the supertwisted $WS_2$ spiral with a twist angle of 16° (Fig. 4c and 4g) and 20° (Fig. 4d and 4h) also show the same oscillating behavior. Another interesting observation is that the onset of SHG decline region changes from 6L to 5L to 4L as the twist angle increases from 13° to 16° to 20°, respectively. Despite being similar thickness, as the twist angle increase, the SHG signal reaches the strongest point faster and start to decline, the interference effect can be ignored compared to structural symmetry and the oscillation period is obviously reduced with the increasing layer number. In addition, the 0° spirals are directly contacted with the substrate (Fig.2c and Fig. 3e); while, there are often a few layers of $WS_2$ existing between the supertwisted $WS_2$ spirals and the substrate (Fig.1a, 1d and Fig. 2g, 2k) since the non-Euclidean surfaces is formed by a few layers of $WS_2$ step edges. As a result, the SHG intensities of the supertwisted spirals are different at the starting point (1L). Overall, depending on the twist angles and layers, the SHG signal can be amplified 12–136 times for the ≈0° spiral,[33,34] 4–92 times for the 13° spiral, 2–45 times for the 16° spirals, and 4-41 times for the 20° spiral, respectively (Fig. S5a, b). Both the maximum and minimum enhancement factors decrease with increasing twist angles. Meanwhile, the SHG intensity of twisted spirals is mainly determined by the twist angle.[33] Herein we evaluate the second-order nonlinear susceptibility of supertwisted $WS_2$ spiral $\chi^{(2)}_{ss-WS_2}$ by compare the SHG intensity between supertwisted $WS_2$ spiral and twisted $WS_2$ at 0° (0.68 nm/V, SI Calculation Method).[33] As shown in Table 1, the maximum of $\chi^{(2)}_{ss-WS_2}$ is reduced from 0.68, 0.46, 0.23, to 0.21 nm/V when the twist angle is increased from 0°, 13°, 16° and 20°.

To explain the oscillating SHG behavior, we further analyzed the effect of inversion symmetry, as show in Fig 5. Here we use a stacked multilayers $WS_2$ model (Fig 3d) to represent the local structural inversion symmetry of supertwisted spirals. Fig. 5a and 5b corresponds to the top view and side view of stacked multilayer $WS_2$ structures. Given $\alpha$=20°, the twist angle between the first two layers is 20° (Fig. 5b), and the twist angle between the first layer and the third layer is 40° (Fig. 5c). They all exhibit broken inversion symmetry, resulting in increased SHG intensity with the increase of layer numbers. In contrast, as the twisted layer increases toward the fourth layer, the twist angle between the fourth layer with the first layer is 60° (the last image of Fig. 5a, b). The inversion symmetry between the first layer and the fourth layer is then restored, resulting a reduced of overall SHG. However, since the fourth layer maintains a twist angle of 40° and 20° relative to the second and the third layers, respectively, the overall SHG intensity would not decrease to zero (Fig. 4h). In addition, Hsu *et al.* found that twist angles of 2°, 16°, 30°, and 37° all resulted in an enhancement of the SHG intensity in the stacked region with respect to monolayer $MoS_2$.[36,错误!未找到引用源。] However, when the twist angle is 54°, the SHG intensity is smaller than that of monolayer $MoS_2$, suggesting that the SHG of twisted bilayer $MoS_2$ is a coherent superposition of SHG fields from individual layers.[错误!未找到引用源。,8] The SHG of a 2H phase hexagonal $WS_2$ spiral was also examined, and the results are shown in Fig. S5 c-f. The twist angle between two screw dislocations is 60° (Fig. S5e), resulting in the recovery of inversion symmetry in hexagonal spiral.[33] Fig. S5f confirms that the SHG signal disappears in even layers. In short, for supertwisted $WS_2$ spiral, if the cumulative twist angle is within 60°, the whole inversion symmetry is broken (Fig. 5c), the nonlinear signal is gradually enhanced with the increase of layer number; when the cumulative twist angle is equal or greater than 60 degrees, the local inversion symmetry is restored because the local specific two layers (Fig. 5c). These specific two layers may suffer from the SHG quenching effect, the nonlinear signal is gradually weakened with the increase of layer numbers and then a periodic oscillation occurs. At the same time, the oscillation frequency of the nonlinear signal is related to the twist angle, a larger twist angles of supertwisted spiral facilitates the production of the SHG intensity with larger oscillation frequency. Therefore, the SHG intensity of supertwisted spirals mainly depends on the inversion symmetry broken (twist angles) of the crystal structure, compared to layer thickness.错误!未找到引用源。-42

## Conclusions

In summary, we found that the atomically supertwisted $WS_2$ spirals with different twist angles can be grown on Euclidean or non-Euclidean surfaces using a one-step water vapor-assisted chemical vapor deposition. The SHG signals in these spirals can be amplified up to 2–136 times depending on their twist angles. The spectroscopic measurements in these spirals suggest a continuous and subtle evolution of atomic configuration. The inversion symmetry is modified as the twist angle changes, leading to a faster increase in SHG intensity of the twisted spirals with smaller twist angles. In a certain range of small twist angle ($\approx 0°$), a maximum SHG enhancement was observed with increasing layer numbers, which originates from the broken inversion symmetry. Due to the recovery of local inversion symmetry, this phenomenon is difficult to be obtained in supertwisted spiral structure with non-zero twist angle. Our work implies that the supertwisted $WS_2$ spirals could provide an attractive platform for the study of light-matter interactions.

**Materials and Methods**

**Water vapor-assisted chemical vapor deposition growth of $WS_2$ spirals**

The atomically supertwisted $WS_2$ spirals were synthesized by water vapor-assisted chemical vapor deposition on a silicon wafer covered with 280 nm $SiO_2$. $WS_2$ powder in a quartz boat was placed at the heating center of a quartz tube in the CVD setup. 0.05–0.1 ml deionized water was placed upstream outside the heating zone as the water vapor source. The substrates were placed about 6–8 cm downstream (700–800 °C) of the quartz tube. High purity argon with flow rate of 80 sccm was introduced into the CVD system for 40 min to discharge the oxygen inside the tube before heating. Then, the furnace was heated to 1150 °C in 35–40 min with an argon flow of 80 sccm and maintained at this temperature for 15–25 min for spiral $WS_2$ growth.

**Material characterizations**

The morphology of spiral $WS_2$ was characterized by optical microscopy (Zeiss Axio Imager A1) and atomic force microscopy (AFM Bruker Multimode 8-HR, AFM probe: Scanasyst-AIR with $f_0$=70 kHz、$k$=0.4 N/m、$T$=650 nm、$L$=115 μm and $W$=25 μm.). SHG measurements were performed using a custom-built confocal microscope in reflection geometry and a Ti: Sapphire laser (Chameleon Ultra) at 800 nm. The laser was vertically incident on the spiral $WS_2$. A 100× microscope objective (Olympus, NA = 0.9) was used to collect reflection light. The SHG signals through a short-pass filter (Thorlabs, cut-off wavelength: 532 nm) were recorded by a charge-coupled device (CCD). The size of the laser spot was 2 μm, and power was 84 $mW/cm^2$.

**ASSOCIATED CONTENT**

Supporting Information

Material characterizations of $WS_2$, including experimental setup for material synthesis, AFM and optical images; additional calculation method of the second-order nonlinear susceptibility $\chi^{(2)}$.

**ACKNOWLEDGMENTS**

The authors express their gratitude to various organizations for their support in this research, including the National Natural Science Foundation of China (Grant No. 52273252, 11804408, 62105232, 52072006, U23A2076), the Fundamental Research Program of Shanxi Province (Grant No. 20210302124027, 20210302124169, 202203021221080).

**Author contributions**

The project was designed and managed by XPF and QZ. The samples were synthesized and characterized by TT. AFM measurements were provided by TT and RJC. SHG measurements were obtained by TT and YY. Interpretation of data was provided by XPF, QZ, QJS, JC, YXK, ZYG, HYW and YYH. XPF and QZ revised the manuscript. All authors reviewed and contributed the revision of the manuscript.

[†]TT, and RJC contributed equally to this work.

**Conflict of interest**

The authors declare no competing interest.

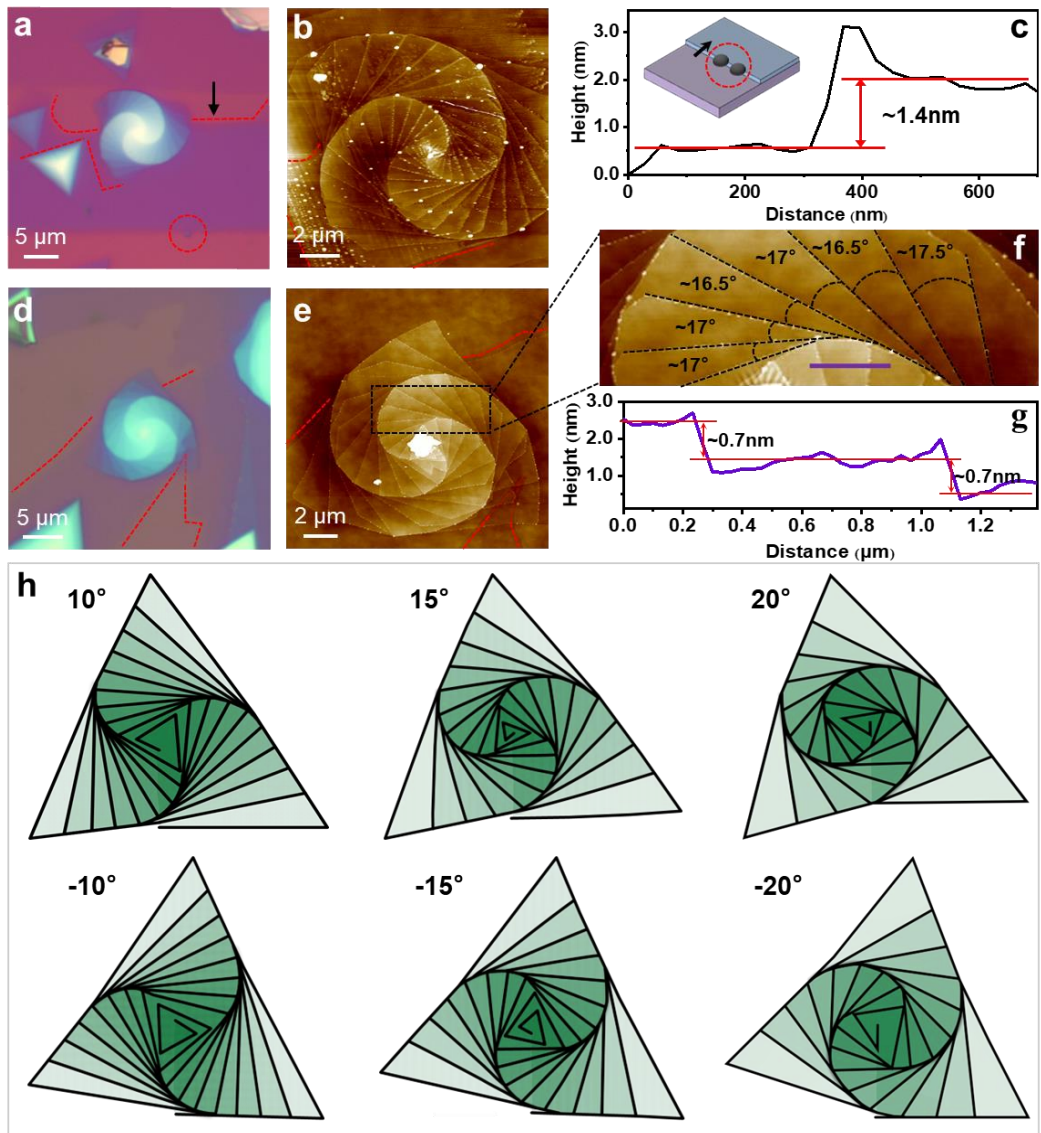

**Figure 1. Atomically Supertwisted WS₂ spiral with different twist angles. a, b** Optical and AFM image of the left-handed supertwisted WS₂ spiral. **c** Line profile marked with a black line in (b) showing that the height of the ribbon was about 1.4 nm. This suggested that twisted spiral WS₂ grows on the ribbon. **d, e** Optical and AFM image of the right-handed supertwisted WS₂. **f** Line profile marked with a purple line in upper showing that the height of each layer was about 0.7 nm. **g** the twist angle was defined by the edges between adjacent layers. **h** Structural model of the left/right-handed supertwisted WS₂ spiral with different twist angles.

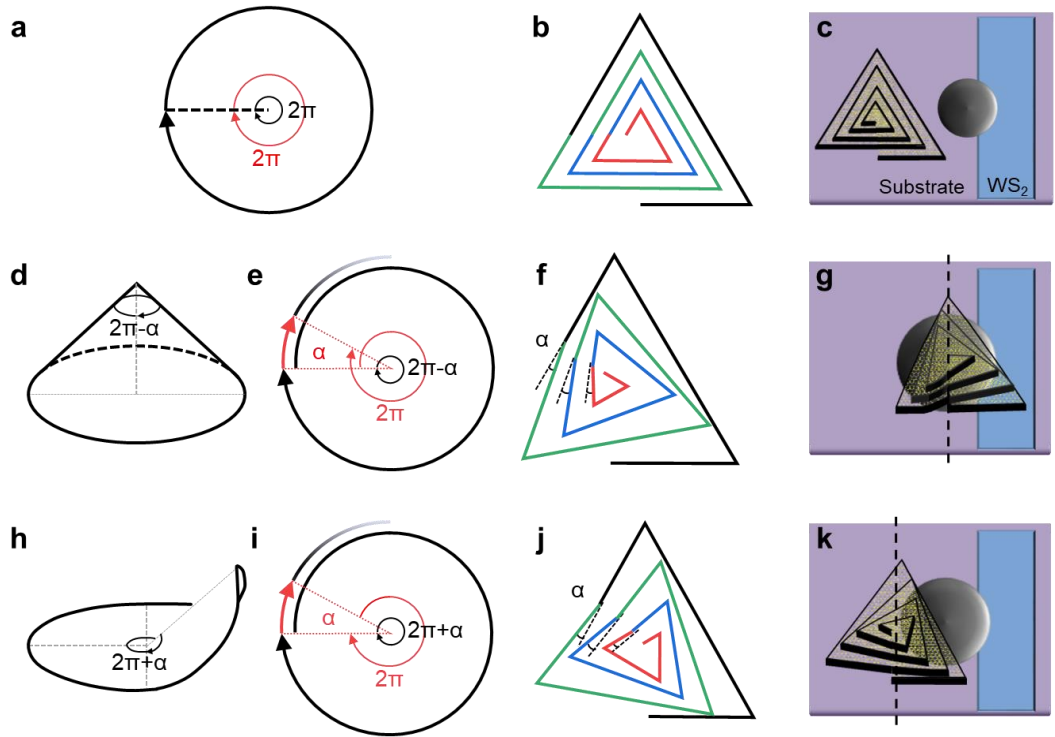

**Figure 2. Formation mechanisms of supertwisted spiral MX$_2$ on different surfaces.**
**a,** $2\pi$ angular period of lattice and substrate with a Euclidean surface. The black circular arrow indicates sample angular period, the red arrow indicates $2\pi$ angular period. **b,** Schematic structure of a right-handed MX$_2$ spiral. Each layer with $2\pi$ angular period is marked with different color. **c,** Schematic growth model of spiral MX$_2$. The purple region represents the SiO$_2$/Si wafer, the blue region represents a WS$_2$ nanoribbon, and the gray circle represents a WO$_x$ nanoparticle formed on the edge of the WS$_2$ nanoribbon. **d, e** $2\pi$-$\alpha$ angular period on a non-Euclidean cone surface, the conical surface is formed due to clipping an angle of $\alpha$. **f,** Schematic crystal structure of the left-hand twisted spiral MX$_2$, with the dislocation center at the apex of the cone (**g**). **h, i** $2\pi$+$\alpha$ angular period on a non-Euclidean "hyperbolic cone" surface, the surface is firmed due to "borrowing" $\alpha$ degree from the next period. **j** Schematic crystal structure of the left-hand twisted spiral MX$_2$, with the dislocation center at the edge of the cone (**k**)

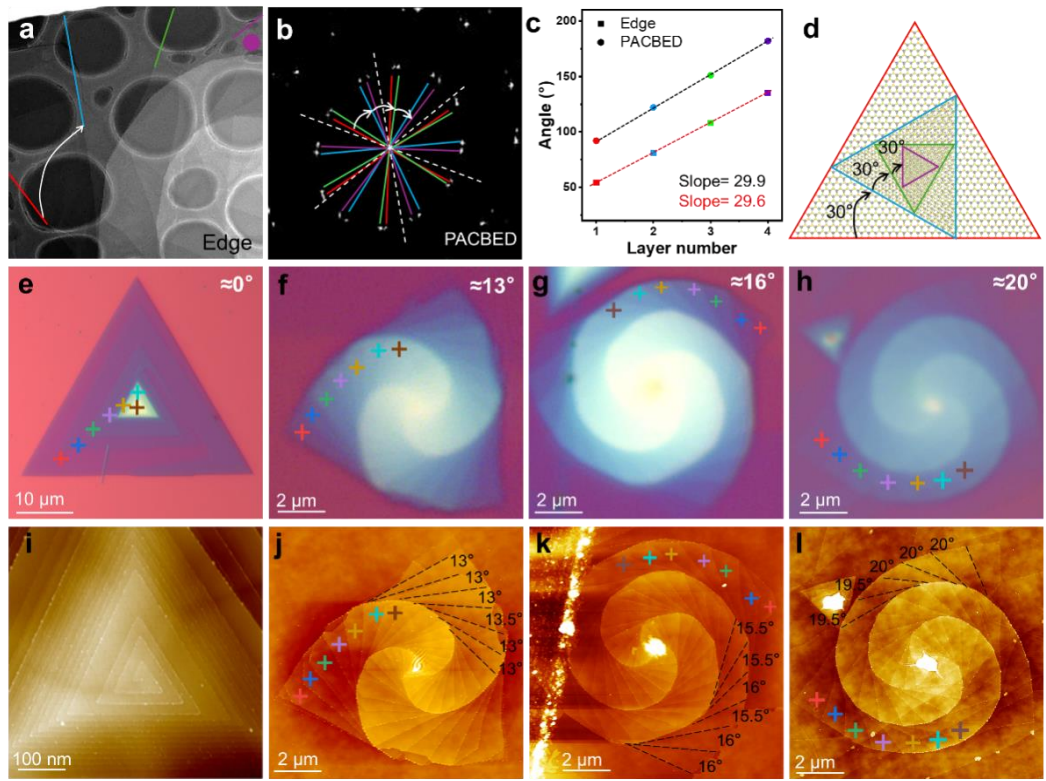

**Figure 3. Characterization of supertwisted WS$_2$ spiral. a** High-angle annular dark-field (HAADF) image of a supertwisted WS$_2$ with edges of different layers highlighted by different colors. **b** the position-averaged convergent beam electron diffraction (PACBED) collected from the purple point region in (a). **c** Twist angles measured by edges and diffraction as functions of the layer number, $\alpha_{Edge}$ =29.9° and $\alpha_{PACBED}$ =29.6°. **d** Schematic diagram of the stacked trilayer WS$_2$ with α=30°. **e-l** Optical images and AFM images of twisted spiral WS$_2$ with twist angle at α≈0° (e, i) α≈13° (f, j), α≈16° (g, k) and α≈20° (h, l), respectively. The colored crosses in these figures show the region where SHG signals are collected.

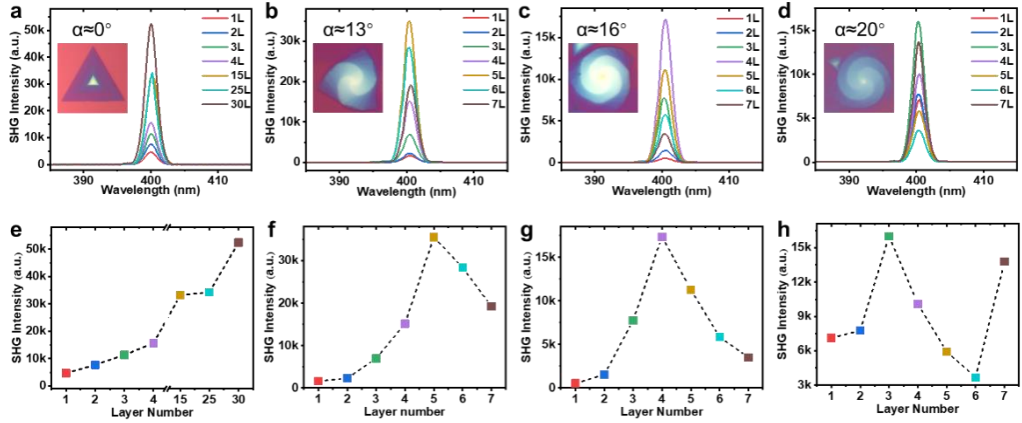

**Figure 4. Angle-dependent SHG of twisted spiral WS₂. a-d** layer-dependent SHG spectra of the WS₂ spiral with twist angle at α≈0° (a), α≈13° (b), α≈16° (c) and α≈20° (d), the insert is an optical picture of the test sample. **e-h** their layer-dependent SHG intensity corresponding to (a-d), respectively. Because a few layers of WS₂ exist between supertwisted WS₂ spirals and substrate, the SHG intensities of supertwisted WS₂ spirals in (f-h) are different at the starting points (1L).

Table 1. The $\chi^{(2)}_{ss-WS_2}$ of supertwisted WS$_2$ spirals with twist angle at $\approx 0°$, 13°, 16° and 20°.

| | Supertwisted WS$_2$ Spirals | | | |
|---|---|---|---|---|
| | $\approx 0°$ | $\approx 13°$ | $\approx 16°$ | $\approx 20°$ |
| $\chi^{(2)}_{ss-WS_2}$ (nm/V) | 0.68 | 0.46 | 0.23 | 0.21 |

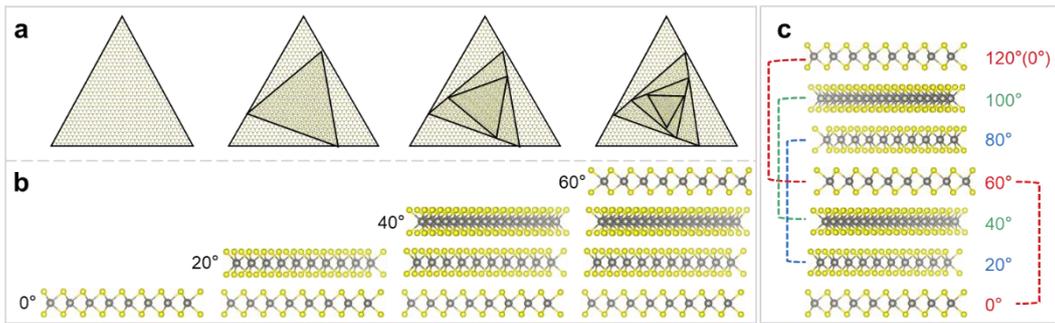

**Figure 5. Inversion symmetry of supertwisted WS₂ spiral. a, b** Schematic diagram of top and side view model of the stacked four-layer WS₂ structure with $\alpha$=20°. **c** the side view of 7 layers stacked multilayers WS₂ with 20° twist angle, the several layers belonging to the same color contour indicates that the local inversion symmetry is restored after rotation of 60°.